\documentclass[aps,prl,reprint,longbibliography,superscriptaddress,twocolumn,floatfix]{revtex4-2}
\usepackage[T1]{fontenc}
\usepackage[utf8]{inputenc}
\usepackage{graphicx}
\usepackage{amsmath}
\usepackage{physics}
\usepackage{xcolor}
\usepackage{txfonts}
\usepackage{times}
\usepackage[pdfpagelayout=TwoPageLeft]{hyperref}
\usepackage{orcidlink}
\usepackage{enumitem}

\definecolor{bggreen}{RGB}{185,230,70}
\definecolor{myblue}{RGB}{0, 40, 140}

\hypersetup{
    unicode=true,          
    pdftitle={Searching for New Fundamental Interactions via Isotopic Shifts in Molecular Lattice Clocks},    
    pdfauthor={},     
    colorlinks=true,       
    linkcolor=myblue,          
    citecolor=myblue,        
    filecolor=myblue,      
    urlcolor=myblue,           
}

\begin{document}
	\title{Searching for New Fundamental Interactions via Isotopic Shifts in Molecular Lattice Clocks}

    \author{E. Tiberi\orcidlink{0000-0001-7168-7194}}
        \altaffiliation{These authors contributed equally.}
        \affiliation{Department of Physics, Columbia University, 538 West 120th Street, New York, NY 10027-5255, USA}
        \affiliation{Martin A. Fisher School of Physics, Brandeis University, MS 057, 415 South Street, Waltham, MA 02453, USA}

    \author{M. Borkowski\,\orcidlink{0000-0003-0236-8100}}
        \altaffiliation{These authors contributed equally.}
        \affiliation{Department of Physics, Columbia University, 538 West 120th Street, New York, NY 10027-5255, USA}
        \affiliation{Van der Waals-Zeeman Institute, Institute of Physics, University of Amsterdam, Science Park 904, 1098 XH Amsterdam, The Netherlands}
        \email{mateusz@cold-molecules.com}
        
    \author{B. Iritani\orcidlink{0000-0002-7911-2755}}
        \affiliation{Department of Physics, Columbia University, 538 West 120th Street, New York, NY 10027-5255, USA}

    \author{R. Moszynski\orcidlink{0009-0008-7669-3751}}
        \affiliation{Quantum Chemistry Laboratory, Department of Chemistry,
        University of Warsaw, Pasteura 1, 02-093 Warsaw, Poland}

    \author{T. Zelevinsky\orcidlink{0000-0003-3682-4901}}
        \email{tanya.zelevinsky@columbia.edu}
        \affiliation{Department of Physics, Columbia University, 538 West 120th Street, New York, NY 10027-5255, USA}
    
	\date{\today}

\begin{abstract}
    Precision measurements with ultracold atoms and molecules are primed to probe beyond-the-Standard Model physics. 
    Isotopologues of homonuclear molecules are a natural testbed for new Yukawa-type mass-dependent forces at nanometer scales, complementing existing mesoscopic-body and neutron scattering experiments. 
    Here we propose using isotopic shift measurements in molecular lattice clocks to constrain these new interactions from new massive scalar particles in the keV$/c^2$ range: the new interaction would impart an extra isotopic shift to molecular levels on top of one predicted by the Standard Model. For the strontium dimer, a Hz-level agreement between experiment and theory could constrain the coupling of the new particles to hadrons by up to an order-of-magnitude over the most stringent existing experiments. 
\end{abstract}
\maketitle

The desire to adequately explain dark matter and the hierarchy problem motivates the search for new gravity-like interactions across many fields of physics. One such hypothetical interaction could emerge from the exchange of a new massive scalar particle resulting in an additional, ``fifth-force'' Yukawa-type interaction between test bodies~\cite{Fayet1996, Adelberger2003, Adelberger2009, Knapen2017, Kamiya2015, Newman2009, Fischbach2012}. Such an interaction is commonly expressed as a non-Newtonian correction of magnitude $\alpha$ to gravitational interaction between masses $m_1$ and $m_2$,
\begin{equation}
    V_5(R) = -\alpha \frac{G m_1 m_2}{R} e^{-R/\lambda},
    \label{eq:yukawa_potential}
\end{equation}
where $G$ and $R$ are the gravitational constant and interbody distance. The range of the non-Newtonian Yukawa interaction $\lambda=\hbar/Mc$ depends on the mass of the hypothesized new particle $M$, and dictates the experimental technique used to search for it. Astronomical observations can constrain $\alpha$ at various
scales~\cite{Newman2009,Desmond2018,Hardy17_StellarBoundsYukawa}.
Torsion balances and microcantilevers are sensitive at the everyday meter- to micrometer distances~\cite{Schwarzschild1986, Klimchitskaya2021, Hoskins1985}. Atomic force microscopy~\cite{Dokou2001, Fischbach2001} and measurements of Casimir interactions~\cite{Mohideen1998, Klimchitskaya2020, Jamell2011} can test gravity at micro- to nanometer scales. Finally, the best laboratory constraints at nano- and picometers originate from neutron scattering experiments~\cite{Kamiya2015, Haddock2018, Nesvizhevsky2008, Heacock2021}.

\begin{figure}[b]
    \includegraphics[width = \columnwidth]{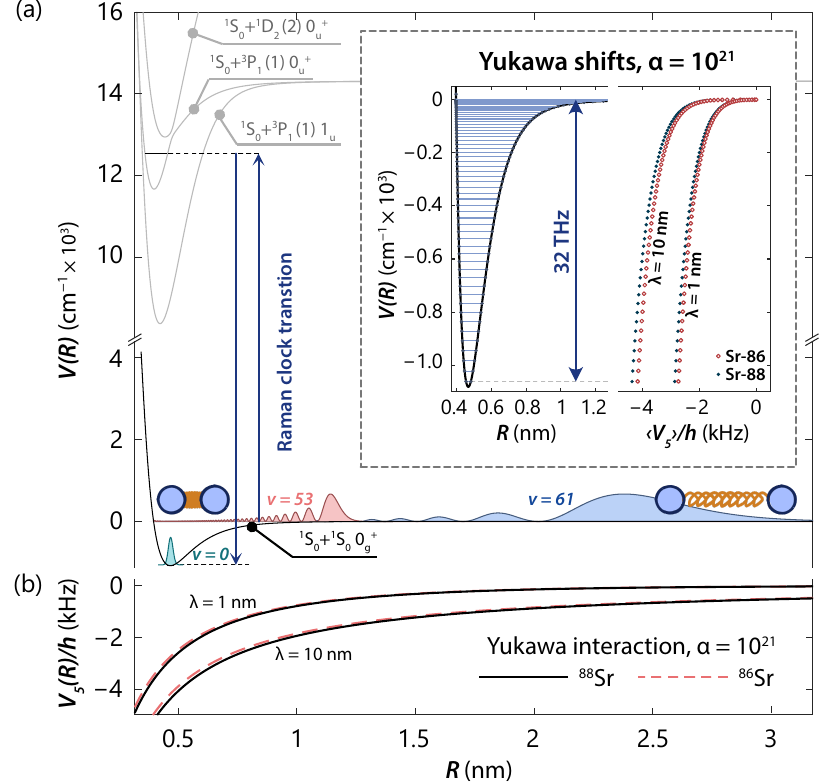}
    \caption{(a) Sr$_2$ potentials~\cite{Stein2010, Skomorowski2012, Borkowski2014} and wavefunctions for selected states $v$. (b) Isotope and range dependence of the Yukawa interaction [Eq.~(\ref{eq:yukawa_potential})] for current limits~\cite{Heacock2021}. Inset: isotope-dependent energy shifts of vibrational levels due to the Yukawa interaction [Eq.~(\ref{eq:perturbation})]. \label{fig:Fig1}}
\end{figure}

The rapid progress in quantum control of atoms and molecules enabled tests of fundamental physics at unprecedented sensitivity~\cite{Safronova2018, Huntemann2014, Godun2014, Wcislo2018, Sanner2019, Kobayashi2019, Lange2021, Figueroa2022, Hur2022, Ono2022, Filzinger2023}. For instance, precise King plot analyses of isotopic shifts in atoms have been used to search for new hadron-lepton interactions \cite{Figueroa2022}. Precision spectroscopy of molecules~\cite{Ubachs2016, Pachomow2017, Bielska2017, Borkowski2017a, Borkowski2018, Zaborowski2020, Kimura2021, Aiello2022, Leung2023, Elmaleh2023, Cozijn2023, Doran2024} offers a complementary new platform for probing new physics. Diatomic molecules are an ideal testbed for new hadron-hadron (rather than hadron-lepton) interactions: their nuclei, each containing $N_1$ and $N_2$ nucleons, become test masses; the extra interatomic potential, $V_5(R) = -N_1 N_2 (g^2/4\pi) e^{-R/\lambda}/{R}$, emerges from the exchange of the new particles between protons and neutrons in the two nuclei~\cite{Yukawa1935, Su1994, Adelberger2009}. The magnitude of the Yukawa correction to gravity depends on the coupling strength $g$ between the new particle field and the nucleons; the conversion factor is $\alpha \approx g^2/(4\pi G u^2)$, where $u$ is the atomic mass unit~\cite{Su1994}. Since molecules have larger natural length scales than atoms, they are also sensitive to longer Yukawa ranges. This corresponds to lighter masses of the hypothetical new particles than those probed by atoms. So far, Salumbides \emph{et al.}~\cite{Salumbides2013} and Germann \emph{et~al.}~\cite{Germann2021} have directly compared transition frequencies in HD$^+$ ions to precise \textit{ab initio} calculations to constrain the coupling constant $\alpha$ in the sub-nm range, while Borkowski \emph{et~al.}~\cite{Borkowski2017, Borkowski2019} exploited the sensitivity of near-threshold states to long-range atomic interactions to search for an extra Yukawa potential at several to tens of nanometers. 

Here, we propose to exploit the mass-squared dependence of the Yukawa interaction by measuring isotopic shifts of molecular clock lines [Fig.~\ref{fig:Fig1}(a)]. The Yukawa potential for $\alpha = 10^{21}$ --- comparable to current constraints at nanometers~\cite{Heacock2021} --- imparts an extra atomic interaction potential of up to several kHz, as shown for Sr$_2$ in Fig.~\ref{fig:Fig1}(b). This additional interaction depends on the isotope and therefore creates an extra isotopic shift. The magnitude of the Yukawa interaction could be then constrained by comparing measured isotopic shifts to theoretical predictions. In this Letter, we first calculate the extra isotopic shift due to the Yukawa interaction, Eq.~(\ref{eq:yukawa_potential}) for the representative Sr$_2$ dimer. Then, we show how one can derive a constraint on the magnitude of the Yukawa interaction and project achievable constraints for a Sr$_2$ molecular clock~\cite{Leung2023}. We demonstrate that our method could yield similar or better constraints than leading neutron-scattering experiments~\cite{Heacock2021}. Finally, we discuss the use of other molecular systems using a simple theoretical model.

By comparing isotopic shifts rather than absolute transition frequencies to \textit{ab initio} theory, we isolate mass-dependent interactions and reduce the burden of accuracy on theory. Producing a meaningful limit on the Yukawa interaction by directly comparing transition frequencies would require calculations that are as accurate as molecular clocks, an unlikely feat for a heavy molecule like Sr$_2$. However, the isotopic shift of a vibrational transition can be theoretically separated from the isotope-independent Born-Oppenheimer potential which can be fitted to accurate molecular spectroscopy data~\cite{Stein2010}. Then a variety of smaller beyond-Born-Oppenheimer effects~\cite{Moszynski2003, Lutz2016, Borkowski2017a} must be calculated \textit{ab initio} to experimental accuracy. The isotope-dependent corrections are on the order of several MHz~\cite{Lutz2016}, and computing them to the few-Hz accuracy of modern molecular clocks~\cite{Leung2023, Iritani2023} requires a theoretical accuracy of $10^{-6}$ which could be within reach. A detailed description of these beyond-Born-Oppenheimer effects is outside the scope of this paper, but calculations for Sr$_2$ are underway.

\begin{figure}[t]
    \centering
    \includegraphics[width = 0.96\columnwidth]{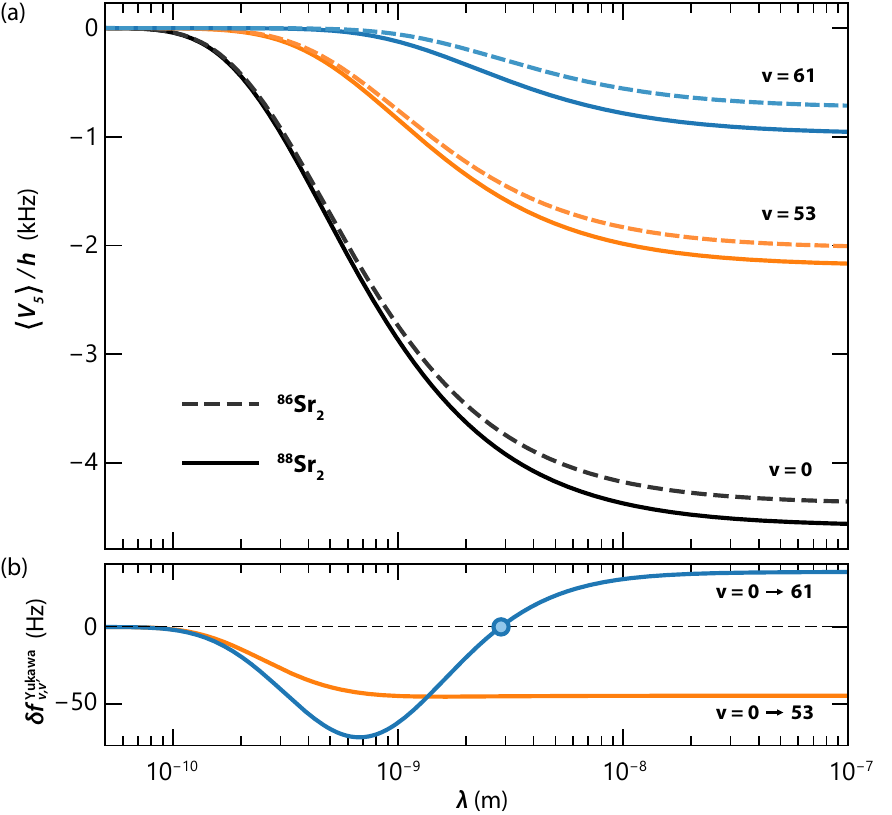}
    \caption{(a) Frequency shift $\langle V_5\rangle/h$ due to a hypothetical Yukawa-type interaction with $\alpha = 10^{21}$ for the rovibrational ground state $v=0$ and selected weakly-bound states $v=53,\,61$. (b) Extra isotopic shift due to Yukawa interaction $\delta f_{v,v'}^{\rm Yukawa}$ for transitions from $v=0$ to weakly bound states $v=53$ and $v=61$. For larger $\lambda$, contributions from very weakly-bound states can overtake that of $v=0$, resulting in a node. 
    \label{fig:Fig2}}
\end{figure}

We first evaluate the sensitivity of molecular transitions to new physics by calculating the extra isotopic shift due to the new Yukawa-type potential, Eq.~(\ref{eq:yukawa_potential}). The Yukawa-type shifts to a given vibrational state are expected to be very small, on the order of a few kHz for the current constraints~\cite{Heacock2021}. We can calculate these shifts using first-order perturbation theory on the numerically calculated vibrational wavefunction of the state $v$, $\Psi_v(R)$. To do so, we solve the vibrational Schr\"odinger equation for the Born-Oppenheimer potential~\cite{Stein2010} using a discrete variable representation (DVR)~\cite{Colbert1992, Tiesinga1998a}. The energy shift for a specific state $v$ is then
\begin{equation}
    \langle V_5\rangle = \int_0^{+\infty} \Psi_v(R) V_5(R) \Psi_v(R) dR. \label{eq:perturbation}
\end{equation}
This energy shift imparts an overall shift in the transition frequency measured between two vibrational states, $v$ and $v'$, 
\begin{equation}
    \delta f_{v,v'} = \frac{1}{h}\left(\langle V_5\rangle_{v} - \langle V_5\rangle_{v'} \right).
\end{equation}
Finally, the extra isotopic shift due to the hypothetical Yukawa interaction --- that is, the extra shift in addition to the expected Standard Model isotopic shift --- between two isotopologues with total atomic mass $A'$ and $A$ is 
\begin{equation}
    \delta f_{v, v'}^{\rm Yukawa} = \delta f_{v,v'}^{A}-\delta f_{v,v'}^{A'}.
    \label{eq:Yuk_shift}
\end{equation}
We note that, unlike in atoms~\cite{Dzuba2022}, the use of perturbation theory is warranted here as the vibrational motion at short ranges is constrained by the repulsive potential wall. Thus, for all physically accessible internuclear separations, the Yukawa interaction is much smaller in magnitude than the Born-Oppenheimer potential.

Since different vibrational wavefunctions in a molecule [Fig.~\ref{fig:Fig1}(a)] probe different internuclear separations, each state has a different sensitivity to the Yukawa interaction. While the rovibrational ground state has a mean internuclear separation $\tilde{R} \approx R_e$, the equilibrium distance which for Sr$_2$ is $0.47$\,nm~\cite{Stein2010}, in weakly-bound states the wavefunction contributions to the matrix element of Eq.~(\ref{eq:perturbation}) are mostly at the outer turning points~\cite{Jones2006}. These can be located as far as several nm, where the Yukawa interaction is substantially weaker [Fig. \ref{fig:Fig1}(b)]. The sensitivity to the Yukawa potential is therefore strongest for deeply-bound states with small internuclear separations and a transition is most sensitive when it spans states that have large differential sensitivities. We expect the best sensitivity to the isotopic dependence of the Yukawa interaction to occur between the rovibrational ground state and weakly-bound states which have the greatest difference in internuclear separation (Fig. \ref{fig:Fig1}, inset). 

Now we move to constrain the magnitude of the Yukawa interaction as a function of the Yukawa range $\lambda$. Figure \ref{fig:Fig2}(a) shows the energy shift due to the Yukawa potential, Eq.~(\ref{eq:perturbation}), for a coupling constant $\alpha = 10^{21}$ and $\lambda$ ranging from 0.1\,nm to 100\,nm. We compute this shift for selected
vibrational states ($v=0$, $v=53$ and $v=61$) of ground-state $^{86}$Sr$_2$ and $^{88}$Sr$_2$ with total angular momentum $J=0$. Even though comparing $^{84}$Sr and $^{88}$Sr could give greater sensitivity to mass-dependent interactions, we pick this pair of isotopes as the higher natural abundance of $^{86}$Sr makes it easier to conduct a clock experiment with short trap loading times. $^{84}$Sr$_2$ molecules have been produced~\cite{Stellmer2012, Ciamei2017} with special loading techniques and their use would further improve the constraints.
Figure \ref{fig:Fig2}(b) shows the extra isotopic shift, $\delta f_{v,v'}^{\rm Yukawa}$ [Eq.~(\ref{eq:Yuk_shift})], for the selected transitions, $v=0\rightarrow61$ and $v=0\rightarrow53$. 
Overall, we find that the extra isotopic shift for a fixed
$\alpha_0 = 10^{21}$ can be on the order of tens of Hz, detectable with the current molecular clock accuracy~\cite{Leung2023}. As expected, deeply-bound states are overall more sensitive to the Yukawa interaction than weakly-bound states. However, at large $\lambda$ and for very weakly bound states -- states where the outer turning points vary significantly between isotopologues -- the contribution from the weakly-bound state can overtake that of the deeply-bound reference state. For this reason, we select $v=53$ as an additional weakly-bound comparison, since this vibrational state offers the largest overall transition energy while still exhibiting a deeply-bound-state-like response to the Yukawa potential. Comparisons of two or more transitions across isotopologues allow us to impose more stringent overall constraints on the Yukawa potential.

\begin{figure}[t]
    \centering
    \includegraphics[width = \columnwidth]{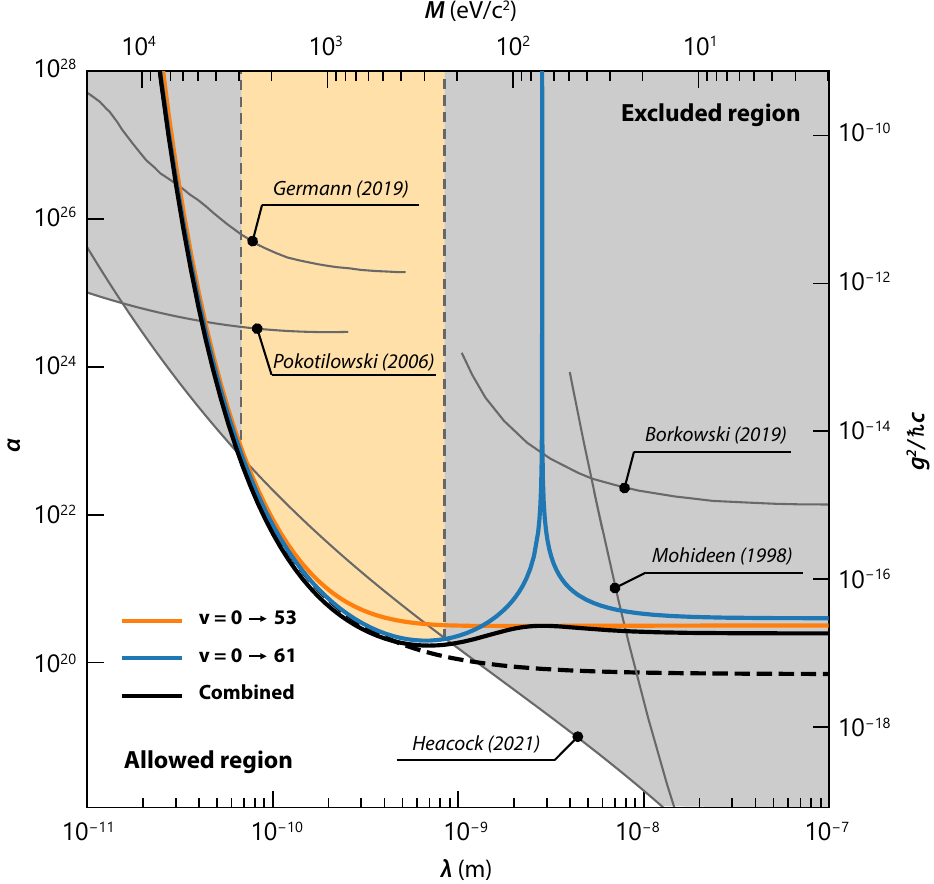}
    \caption{Projected constraints (95\% CI) on $\alpha$ from an isotopic shift measurement of the $v=0\rightarrow 61$ (blue) and $v=0\rightarrow 53$ (orange) transitions in $^{88}$Sr$_2$ and $^{86}$Sr$_2$, as a function of $\lambda$. The combined constraint is shown in black. Previous limits (gray) are set by neutron scattering~\cite{Pokotilovski2006, Heacock2021} and Casimir force measurements~\cite{Mohideen1998}. Other molecule-based methods (Yb$_2$~\cite{Borkowski2019} and HD$^+$~\cite{Germann2021}) are also shown. For completeness, we also show the constraints in terms of particle mass $M$ and squared coupling strength $g^2$. Even for current molecular clock accuracy~\cite{Leung2023} our method projects an order-of-magnitude improvement over leading laboratory constraints~\cite{Heacock2021} for Yukawa forces due to new particles in the keV range. The dashed line represents an analytical approximation to the $^{86}$Sr-$^{88}$Sr constraint [Eq.~(\ref{eq:approximate_limit})]}. 
    \label{fig:Fig3}
\end{figure}

To illustrate the potential of our method for constraining hypothetical Yukawa-type interactions, we construct an exclusion plot (Fig. \ref{fig:Fig3}) which compares our projection against available state-of-the-art laboratory constraints. Consider the following scenario: an isotopic shift is determined for a transition by independently measuring its absolute frequency for each isotope and subtracting. Assuming that both measurements have an accuracy $u(f)=5.1$~Hz~\cite{Leung2023} and that the uncertainty of the theory is insignificant, the combined error bar of the isotopic shift is $\sqrt{2} \times 5.1$\,Hz.  Note that the experimental accuracy has a high potential for improvement.

The experimentally measured isotopic shift, $\delta f_{v, v'}^{\rm exp}$, can be understood as a sum of the Standard Model isotopic shift $\delta f_{v, v'}^{\rm SM}$ and the hypothetical Yukawa contribution $\delta f_{v,v'}^{\rm Yukawa}$, Eq. (\ref{eq:Yuk_shift}). Any difference between the experiment and the Standard Model would be interpreted as an indirect measure of the new Yukawa interaction. Using our computed extra isotopic shift [Fig. \ref{fig:Fig2}(b)], the magnitude of the new interaction $\alpha$ and its uncertainty $u(\alpha)$ can be obtained through a simple proportion:
\begin{equation}
    \frac{\alpha}{\alpha_0} = \frac{\delta f_{v, v'}^{\rm exp}-\delta f_{v, v'}^{\rm SM}}{\delta f_{v,v'}^{\rm Yukawa}(\alpha_0)}
\end{equation}
where $\delta f_{v,v'}^{\rm Yukawa}(\alpha_0)$ is the isotopic shift for $\alpha_0 = 10^{21}$. 

While $\alpha$ determined from experiment can, in principle, be negative, since $\alpha$ is proportional to $g^2$ it is customary to constrain $\alpha$ to be positive, corresponding to an attractive interaction~\cite{Feldman1998, Kamiya2015}. To correct for this, we employ the Feldman-Cousins approach~\cite{Feldman1998} to construct a 95\% confidence limit for $\alpha$, which is $\bar \alpha = f(x_0)\times u(\alpha)$. The multiplicative factor $f(x_0)$ is the upper bound of the confidence interval for the reduced variable $x_0 = \alpha / u(\alpha)$ and varies from $0.62$ for $x_0 = -1$ (i.e., $-1\sigma$ difference between
experiment and theory) to $2.96$ for $x_0 = +1$ (i.e., $+1\sigma$ difference). For our projected constraints we use the representative center case of $x_0 = 0$ and $f(x_0) = 1.96$, which corresponds to full agreement between theory and experiment, and a null result on $\alpha$.

Figure~\ref{fig:Fig3} shows the projected constraint set by two transitions, $v=0\rightarrow61$ and $v=0\rightarrow53$. The projected constraints would be competitive with current limits set by state-of-the-art methods based on neutron scattering~\cite{Pokotilovski2006, Heacock2021} and measurements of Casimir-Polder forces~\cite{Mohideen1998}. Our method would also complement other molecular experiments~\cite{Ubachs2016, Germann2021, Borkowski2019} by filling the gap in Yukawa ranges between the approach based on the very small HD$^+$ molecule ($\lambda \approx 0.1$\,nm) and one based on weakly bound molecules ($\lambda \approx 10\,$nm). Most notably, we project up to an order-of-magnitude improvement on the leading constraint~\cite{Heacock2021} in the 0.1 to 1\,nm range which corresponds to new massive particles in the keV range. This is also the natural length scale for the Sr$_2$ molecule whose equilibrium distance is $0.47$\,nm.

We note that constraints from individual transitions to very weakly-bound states can have resonancelike behaviors in $\lambda$ for ranges that correspond to zero-crossings in the extra isotopic shift (see Figs. \ref{fig:Fig2}(b) and \ref{fig:Fig3}). This effect can be compensated by measuring isotopic shifts for two (or more) transitions, $A$ and $B$. For each transition, we could derive values $\alpha_A$ and $\alpha_B$ and a combined coupling strength, $\alpha_C$, using a simple weighted average with a corresponding uncertainty $u(\alpha_{C})=\left( 1/u^2(\alpha_A) + 1/u^2(\alpha_B)\right)^{-1/2}$. Again assuming a null result, the 95\% confidence limit on $\alpha_{C}$ using Feldman-Cousins~\cite{Feldman1998} is $\bar \alpha_{C} = 1.96 \times u(\alpha_{C})$. In this way, we can effectively remove the resonancelike behavior of the $v=0\rightarrow 61$ constraint.

\begin{figure}[t]
    \centering
    \includegraphics[width = \columnwidth]{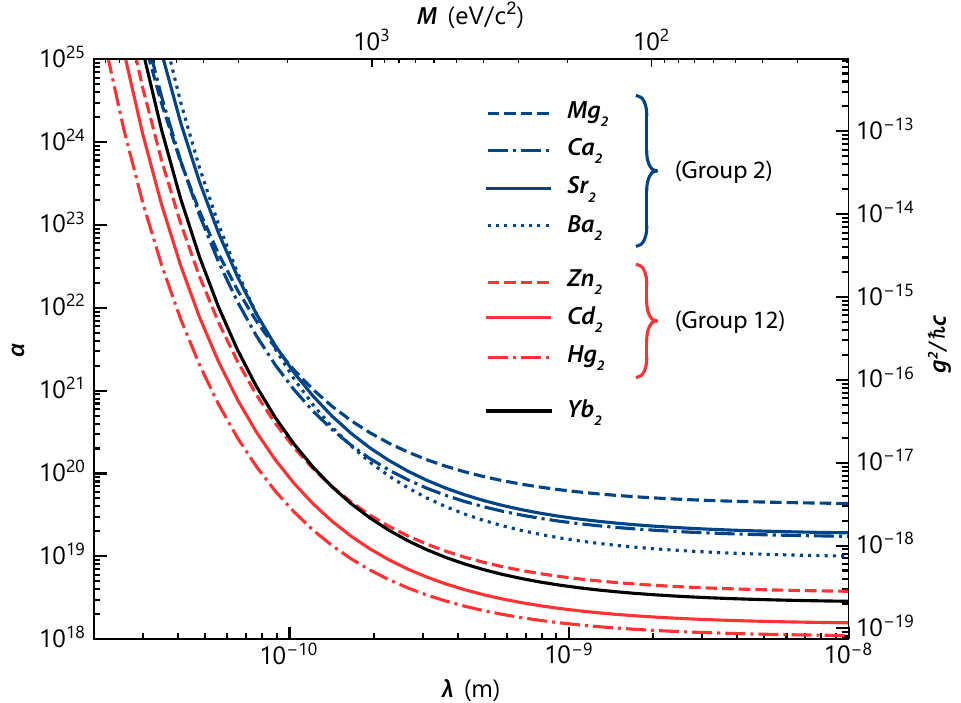}
    \caption{
        Projected limits achievable with different two-valence-electron species using Eq.~(\ref{eq:approximate_limit}) with parameters from Table~\ref{table:all_dimers}. Here we consider several atomic species from Group 2 (blue), Group 12 (red), and the two-valence-electron lanthanide, ytterbium (black). 
        \label{fig:Figure4}
    }
\end{figure}

Lastly, we wish to compare the achievable constraints across different molecular species. Instead of repeating the numerical approach above for each system, we instead use an approximate analytic model. This approximation assumes transitions with the largest sensitivity to a Yukawa-type potential, that is transitions that span the entire depth of the potential. The majority of the sensitivity to the Yukawa potential in these transitions stems from the deeply-bound state, whose wavefunction is localized around the equilibrium distance $R=R_e$. By ignoring contributions from the weakly-bound states, we can approximate the extra isotopic shift as
\begin{equation}
    \delta f_{v,v'}^{\rm Yukawa} \approx \alpha G \frac{m_1^2-m_2^2}{R_e}e^{-R_e/\lambda}/h
\end{equation}
where $m_1$ and $m_2$ are the isotopologue masses.
For transitions between the rovibrational ground state and a near-threshold state, the frequency of the transition, $f$, is close to the dissociation energy of the molecule $D_e$. We can, therefore, write $\alpha$ in terms of properties of the molecular species, $R_e$ and $D_e$, and the relative uncertainty of the experiment, $u(f)/f$:
\begin{equation}
    \log_{10} \alpha 
        \approx f_{\rm mol} 
        + \log_{10} \frac{u(f)}{f}
        + 0.43 \frac{R_e}{\lambda}\, ,
        \label{eq:approximate_limit}
\end{equation}
where the molecular-species-specific factor
\begin{equation}
    f_{\rm mol} = 
        \log_{10} \left[\frac{1.96}{\sqrt{2}} \frac{R_e D_e}{G \bar{m}\, {\delta m}}\right]
        \label{eq:sensitivity_limit}
\end{equation}
depends only on the specific characteristics of the molecule and not on the experimental details. Here we use $\bar m = (m_1+m_2)/2$ and $\delta m = |m_1-m_2|$ and apply $f(x_0)=1.96$ corresponding to a 95\% confidence limit for a null result~\cite{Feldman1998}. 

A cursory look at this approximation reveals that at large Yukawa ranges $\lambda \to \infty$ the constraint on $\alpha$ becomes $\log_{10}\alpha \to \log_{10} [u(f)/f] + f_{\rm mol}$. Ultimately, we wish to find ways to minimize the constraint set on $\alpha$, which can be achieved by either preferentially selecting a molecular species with the lowest $f_{\rm mol}$ or improving experimental and theoretical accuracy. The sensitivity factor, $f_{\rm mol}$, depends on the equilibrium distance $R_e$, potential depth $D_e$ and the variety of available stable isotopologues. In general, heavier species (large $\bar m$), or species with a wider array of isotopologues (large $\delta m$), are more sensitive to the
Yukawa interaction. When comparing performance across species, we consider the lightest and heaviest bosonic (with zero nuclear spin) isotopes available to calculate $f_{\rm mol}$. 

To validate Eq.~(\ref{eq:approximate_limit}), we compare it to our numerical calculation for isotopic shifts between $^{86}$Sr$_2$ and $^{88}$Sr$_2$, assuming $\log_{10}[u(f)/f] = -12.8$~\cite{Leung2023} (dashed line in Fig.~\ref{fig:Fig3}). We find that it agrees semi-quantitatively, in particular at short ranges where the contribution of the deeply-bound state dominates. However, Eq.~(\ref{eq:approximate_limit}) explicitly ignores the residual sensitivity of the weakly-bound state, which would ordinarily reduce the sensitivity of the entire transition, in particular at larger Yukawa ranges $\lambda \gtrapprox 1$\,nm. While we can reasonably reproduce the overall behavior of the constraint, this approximation must be considered optimistic and we use it solely to compare the projected performance of various two-electron species. 

\begin{table}[t]
    \caption{
        Properties of selected spin-singlet homonuclear dimers composed of two-valence-electron atoms. Here $A_{\rm min}$ and $A_{\rm max}$ correspond to the number of nucleons of the lightest and heaviest stable bosonic isotopes. The potential depth $D_e$ and equilibrium distance $R_e$ are taken from literature (``Ref.''). The last column lists the ``molecular sensitivity factor'' $f_{\rm mol}$ [Eq.~(\ref{eq:sensitivity_limit})].
        \label{table:all_dimers}
    }
  \begin{ruledtabular}
  \begin{tabular}{lrrrrcr}
    Species &     $A_{\rm min}$ &     $A_{\rm max}$ &     $D_e$ (cm$^{-1}$) &     $R_e$ (nm) &     Ref. &     $f_{\rm mol}$ \\
    \hline
    Mg$_2$ &     24 &     26 &     430.3 &     0.39 &     \cite{Knockel2013} &     32.42 \\
    Ca$_2$ &     40 &     48 &     1102.1 &     0.43 &     \cite{Allard2002} &     32.02 \\
    Sr$_2$ &     84 &     88 &     1081.6 &     0.47 &     \cite{Stein2010} &     32.06 \\
    Ba$_2$ &     132 &     138 &     1199.0 &     0.52 &     \cite{Li2011} &     31.78 \\
    Zn$_2$ &     64 &     70 &     279.1 &     0.42 &     \cite{Czajkowski1999} &     31.36 \\
    Cd$_2$ &     106 &     116 &     330.5 &     0.41 &     \cite{Czajkowski1999} &     30.97 \\
    Hg$_2$ &     196 &     204 &     379.5 &     0.36 &     \cite{Krosnicki2015} &     30.82 \\
    Yb$_2$ &     168 &     176 &     658.0 &     0.46 &     \cite{Visentin2021} &     31.23 \\
  \end{tabular}
  \end{ruledtabular}
  \label{tab:results}
\end{table}

Table~\ref{table:all_dimers} lists the molecular properties of a number of homonuclear dimers composed of two-valence-electron species that, similar to Sr, are attainable at ultracold temperatures and allow precision spectroscopy~\cite{Kulosa2015, Oates2000, Pachomow2017, De2009, Dzuba2019, Tyumenev2016, Yamanaka2015, Yamaguchi2019, McGrew2018, Kato2012}. The molecular sensitivity factors, $f_{\rm mol}$, for all the considered species are all similar, and would lead to constraints on $\alpha$ within two orders of magnitude of each other. The equilibrium distances for all the molecules are also similar, in the range of $0.36\,$nm for Hg$_2$ to $0.52\,$nm for Ba$_2$. As a result, the limits on $\alpha$ all follow a similar trend and begin to increase for $\lambda$ between $0.1$~nm and $1$~nm (see Fig.~\ref{fig:Figure4}). 

As anticipated, species that are heavier and species with more isotopologues have an overall better sensitivity. This result suggests that heavier species, like Yb$_2$ and Hg$_2$, would be good candidates for constraining new Yukawa forces by precise measurements of isotopic shifts. Yet, this size advantage has to be weighed against the exponentially increasing complexity of \emph{ab initio} calculations for heavier species. As such, the Ca$_2$ molecule presents a particularly attractive option; though lighter than Sr$_2$, Ca$_2$ has nearly the same sensitivity -- mostly due to its wider array of isotopologues -- and is far more tractable theoretically by quantum chemistry methods because it has fewer electrons: 20 per atom as opposed to 38 in Sr.

In conclusion, precision measurements in ultracold spin-singlet molecules is a fruitful avenue to search for new gravity-like interactions~\cite{Fayet1996, Salumbides2013, Germann2021, Borkowski2019}. Here, we have proposed a new method for constraining hypothetical Yukawa-type interactions that relies on comparing isotopic shifts of molecular vibrational transitions to theoretical predictions. Contrary to direct comparison of \emph{ab initio} calculations to experimental molecular spectroscopy in a single isotopologue, our method places far less stringent requirements -- by several orders of magnitude -- on the accuracy of the \textit{ab initio} calculations. A simulation carried out in Sr$_2$ reveals that the achievable limit on the new Yukawa force in the sub-nanometer range could compete with and surpass the state-of-the-art constraints obtained from neutron scattering experiments~\cite{Heacock2021}, if theory can explain isotopic shifts to within Hz-level experimental accuracy~\cite{Leung2023}. A theoretical description of the isotopic shift requires \emph{ab initio} calculations of effects beyond the Born-Oppenheimer approximation, including the simple nuclear mass and volume corrections~\cite{Lutz2016}, as well as the relativistic and QED corrections~\cite{Moszynski2003}. Calculations for Sr$_2$ are currently in progress. We also investigated the potential use of this method for other molecules. In particular, Ca$_2$ can be just as sensitive to the hypothetical fifth force as Sr$_2$ thanks to its wider range of isotopologues while being more tractable theoretically.

\begin{acknowledgments}
    This work was supported by NSF grant PHY-1911959, AFOSR MURI FA9550-21-1-0069, the Brown Science Foundation, the Chu Family Foundation, and the Polish National Science Centre (NCN) grant 2017/25/B/02698.  M.~B. was partially funded by the Polish National Agency for Academic Exchange within the Bekker Programme, project PPN/BEK/2020/1/00306/U/00001.
\end{acknowledgments}

\bibliography{library}

\end{document}